\documentclass[aps,prd,10pt,notitlepage,nofootinbib,superscriptaddress,showkeys]{revtex4}
\usepackage{amstext,amsmath,amssymb,amsfonts,bbm}
\usepackage[latin1]{inputenc}
\usepackage{graphicx}
\usepackage{hyperref}
\usepackage{epstopdf}
\usepackage{amsthm}
\usepackage{pstricks}
\usepackage{tocvsec2}

\usepackage{color}

\def\beq{\begin{equation}}
\def\be{\begin{equation}}
\def\ee{\end{equation}}
\def\bes{\begin{eqnarray}}
\def\ees{\end{eqnarray}}

\DeclareMathOperator{\im}{Im}
\DeclareMathOperator{\Ad}{Ad}
\DeclareMathOperator{\Hom}{Hom}
\DeclareMathOperator{\rk}{rk}

\DeclareMathOperator{\SU}{SU}
\DeclareMathOperator{\SO}{SO}

\DeclareMathOperator{\U}{U}

\newcommand{\alg}{\mathfrak{g}}
\newcommand{\su}{\mathfrak{su}}

\renewcommand{\u}{\mathfrak{u}}

\newcommand{\unit}{\mathbbm{1}}

\def\bra{\langle}
\def\ket{\rangle}
\def\f{\frac}

\def\mone{^{-1}}

\def\pp{\partial}

\def\Z{{\mathbbm Z}}

\def\calA(\Gamma,G){{\mathcal A}}

\def\calF{{\mathcal F}}

\def\calM{{\mathcal M}}

\def\calO{{\mathcal O}}

\def\calZ{{\mathcal Z}}

\def\u{\underset}
\def\k{K_{\tau}}

\theoremstyle{definition}
\theoremstyle{definition}
\theoremstyle{definition}
\theoremstyle{definition}
\theoremstyle{definition}
\theoremstyle{definition}

\begin{document}
\maxtocdepth{subsection}
\pagestyle{plain}

\title{\large \bf Bubble divergences: sorting out topology from cell structure}

\author{{\bf Valentin Bonzom}}\email{vbonzom@perimeterinstitute.ca}
\affiliation{Perimeter Institute for Theoretical Physics, 31 Caroline St. N, ON N2L 2Y5, Waterloo, Canada}

\author{{\bf Matteo Smerlak}}\email{smerlak@cpt.univ-mrs.fr}
\affiliation{Centre de Physique Th\'eorique, Campus de Luminy, Case 907, 13288 Marseille Cedex 09 France}

\date{\small\today}

\begin{abstract}\noindent
We conclude our analysis of bubble divergences in the flat spinfoam model. In \href{http://arxiv.org/pdf/1008.1476}{[arXiv:1008.1476]} we showed that the divergence degree of an arbitrary two-complex $\Gamma$ can be evaluated \emph{exactly} by means of twisted cohomology. Here, we specialize this result to the case where $\Gamma$ is the two-skeleton of the cell decomposition of a pseudomanifold, and sharpen it with a careful analysis of the cellular and topological structures involved. Moreover, we explain in detail how this approach reproduces all the previous powercounting results for the Boulatov-Ooguri (colored) tensor models, and sheds light on algebraic-topological aspects of Gurau's $1/N$ expansion. 

\end{abstract}

\maketitle


\section{Introduction}

This paper is the third and last of a series \cite{Bonzom:2010fk,Bonzom:2010uq} concerned with the divergences of the \emph{flat spinfoam model}, defined by the formal integral

\be \label{partition}
\mathcal{Z}(\Gamma,G):=\int_{\mathcal{A}(\Gamma,G)}dA\ \prod_{f} \delta\bigl(H_f(A)\bigr),
\ee
where $\Gamma$ is a $2$-complex,  $G$ a compact Lie group, $A\in\mathcal{A}(\Gamma,G)$ a discrete $G$-connection on $\Gamma$ and $H_f(A)$ the holonomy of $A$ around the face $f$ of $\Gamma$ (precise definitions will be given in the next section). This problem is important from three perspectives.
\medskip\begin{itemize}
\item
\emph{The viability of the spinfoam approach to quantum gravity} \cite{Rovelli:2004tv}. The expression \eqref{partition} is the prominent instance of a \emph{spinfoam model}. It includes the Ponzano-Regge \cite{PR} and Ooguri models \cite{Ooguri:1992eb} in three and four dimensions, and all the models discussed in the literature are modifications thereof. Understanding the (mathematical as well as physical) nature of its divergences is therefore required to assess the whole spinfoam approach to quantum gravity.

\item
\emph{The renormalization program in group field theory.} A closely related framework for the quantization of gravity is provided the \emph{group field theory}  introduced by Boulatov \cite{Boulatov:1992vp} and reviewed in \cite{Freidel:2005qe,Oriti:2006se}. From this perspective, the expression \eqref{partition} is a typical (vacuum) Feynman amplitude, and the first question to address is its scaling behaviour. Does \eqref{partition} define a renormalizable quantum field theory in some generalized sense \cite{Rivasseau:2007ab}? In parallel to general powercounting considerations, some results beyond the leading divergent order have been recently obtained in \cite{2ptboulatov}, showing the need for both mass and wave-function renormalizations in the Boulatov-Ooguri tensor model.

\item
\emph{Quantum topology}. It has long been hypothesized that \eqref{partition} should define a topological invariant of manifolds. (It was in fact the first state-sum model ever written, and inspired Turaev and Viro in defining their invariant.) Giving it a proper mathematical definition is considered an interesting problem \emph{per se} \cite{Barrett:2009ys}. A special case of \eqref{partition}, corresponding to knot exteriors, has been addressed recently in \cite{frohman-dubois-torsion} and shown to display an interesting connection with the Dubois prescription for the Reidemeister torsion of knot exteriors \cite{these-dubois}.

\end{itemize}

From a mathematical perspective, the expression \eqref{partition} is ill-defined for the same reason that a field-theoretic Feynman ampitude is ill-defined: it involves products of distributions. To unravel the structure hidden by these products, the strategy is standard: introduce a cutoff $\Lambda$ and a smooth approximation $\delta_\Lambda$ of the Dirac $\delta$, and study the behaviour of

\be
\mathcal{Z}_\Lambda(\Gamma,G):=\int_{\mathcal{A}(\Gamma,G)}dA\ \prod_{f} \delta_\Lambda\bigl(H_f(A)\bigr)
\ee
as $\Lambda\rightarrow\infty$. If it exists, the number $\Omega(\Gamma,G)$ such that
\be\label{regularized}
\mathcal{Z}'(\Gamma,G):=\underset{\Lambda\rightarrow\infty}{\lim}\ \Lambda^{-\Omega(\Gamma,G)}\mathcal{Z}_\Lambda(\Gamma,G)
\ee
is finite and non-zero can be called the \emph{divergence degree} of $\Gamma$. This paper, like \cite{Bonzom:2010fk,Bonzom:2010uq}, deals with the evaluation of this number.

Since Perez and Rovelli's work \cite{Perez:2000fs}, it has usually been assumed that $\Omega(\Gamma,G)$ is related to the number of ``closed surfaces" or ``bubbles" within $\Gamma$ -- hence the nickname ``bubble divergences". In \cite{Bonzom:2010fk}, we discussed the conditions under which this intuition is valid: when $\Gamma$ is simply connected or when $G$ is Abelian, one finds indeed that $\Omega(\Gamma,G)=(\dim G)b_2(\Gamma)$, with $b_2(\Gamma)$ the second Betti number of $\Gamma$ (i.e. the number of independent $2$-cycles in $\Gamma$). In that note, however, we emphasized that such a straightforward relationship between the combinatorics of $\Gamma$ and its divergence degree does \emph{not} hold in general. In particular, it turns out that $\Omega(\Gamma,G)$ is usually not an integer multiple of $\dim G$. To estimate it in these cases, we developed in \cite{Bonzom:2010uq} a more general cohomological framework, known as \emph{twisted cohomology}, for the powercounting of the flat spinfoam model. We showed that, except for the presence of certain singularities in the moduli space of flat discrete connections on $\Gamma$, for which a general treatment remains to be found, the general formula for $\Omega(\Gamma,G)$ is given by the \emph{second twisted Betti number} of $\Gamma$.

Although rather general, our method left unanswered an important question. In most cases, indeed, $\Gamma$ arises as the $2$-skeleton of some cell decomposition $\Delta_M$ of a pseudomanifold $M$ \footnote{Most considerations in the present paper actually apply more generally to cell complexes $\Delta$ whose 2-skeleton is $\Gamma$. However, the case of pseudo-manifolds is the natural choice in the framework of group field theories and is particularly interesting from the view of quantum invariants.}. In the spinfoam formalism, one often starts with a triangulation $T_M$ of a manifold $M$, and $\Delta_M$ is the dual cell decomposition of $M$. From the group field theory perspective, on the other hand, $\Gamma$ is a generalized Feynman diagram, which defines a pattern for gluing simplices along their codimension $1$ faces, like in matrix models. The quotient space is a pseudomanifold $M$ \cite{DePietri:2000ii,Smerlak:2011rd} which, under certain further assumptions \cite{Gurau:2009tw}, is naturally equipped with a ``singular triangulation" \cite{matveev}; $\Delta_M$ is then the cell decomposition of $M$ dual to this ``singular triangulation". What is the relationship between the divergence degree $\Omega(\Gamma,G)$ and the topology of $M$?

The main result of this paper consists in showing that $\Omega(\Gamma,G)$ depends on the topology of $M$, \emph{but also on the particular cell decomposition} $\Delta_M$, according to
\be\label{mainresult}
\Omega(\Gamma,G)=I(M,G)+\omega(\Delta_M,G).
\ee
Here, $I(M,G)$ is a topological invariant of $M$, depending only on its fundamental group and Euler characteristic, and $\omega(\Delta_M,G)$ is a simple function of the numbers of $k$-cells in $\Delta_M$, with $0\leq k\leq\dim M$.

This decomposition, besides clarifying the respective r\^oles played by the topology of $M$ and the cellular structure $\Delta_M$ in the powercounting of \eqref{partition}, has the merit of allowing for a precise comparison with various other results, in particular those of Freidel and Louapre \cite{Freidel:2004vi,Freidel:2004nb}, Freidel \emph{et al.} \cite{Freidel:2009hd}, Magnen \emph{et al.} \cite{Magnen:2009at} and Ben Geloun \emph{et al.} \cite{Geloun:2010nw}. In a nutshell: all of them are \emph{particular cases} of \eqref{mainresult}. A substantial fraction of this paper will be devoted to explaining this point.

One of the aspects which had so far made this comparison difficult is the usage of a somewhat different language in our papers \cite{Bonzom:2010fk,Bonzom:2010uq} and in \cite{Gurau:2009tw,Geloun:2010nw}. While we use the standard cellular (co)homology of two-dimensional complexes, Gurau, Rivasseau and their collaborators have championed a combinatorial approach based on \emph{colored graphs}. Indeed, Gurau has showed that the latter are naturally the Feynman graphs of a modified group field theory coined ``colored group field theory" \cite{Gurau:2009tw}. Various arguments have been given in favor this framework: it avoids certain topological singularities \cite{Gurau:2010nd},\footnote{It is claimed in \cite{Gurau:2010nd} that only colored group field theory ensures the avoidance of more singular spaces than pseudomanifolds. One of us (M.S.) has disputed this claim \cite{Smerlak:2011rd}.} has a well-defined ``bubble homology"  \cite{Gurau:2009tw} and exhaustive sequences of cuts in the sense of \cite{Magnen:2009at}; it yields a $1/N$ expansion mimicking the 't Hooft planar limit of matrix models \cite{Gurau:2010ba,Gurau:2011aq}; and it permits the identification of diffeomorphims in group field theories \cite{Baratin:2011tg}. All these results are interesting, and our approach based on two-complexes could rightfully be considered dull if it could not handle them. However, \emph{it can}: there is a precise relationship between colored graphs and two-complexes, and their natural (co)homologies. Again, our approach is not different from the one based on colored graph: it is simply more general.


The plan of the paper is as follows. In sec. \ref{reminder}, we review our derivation of the divergence degree $\Omega(\Gamma,G)$ from twisted cohomology and derive its formulation \eqref{mainresult}. Sec. \ref{comparison} is devoted to the comparison of this formula with the other results discussed in the literature. In sec. \ref{examples}, we illustrate \eqref{mainresult} with several three-dimensional examples. Sec. \eqref{conclusion} presents our conclusion.

\section{Powercounting from twisted cohomology}\label{reminder}
Let us begin by reviewing our analysis of bubble divergences in terms of twisted cohomology.

\subsection{Discrete connections}

Let $G$ be a compact Lie group, whose Lie algebra $\alg$ is equipped with an invariant inner product, and $\Gamma$ a two-dimensional piecewise linear CW complex (see Appendix A for the definition of a CW complex). Denote $V, E, F$ the number of vertices, edges and faces respectively. The notation $\Gamma_i$ will be used for the set of $i$-cells of the complex. 

A \emph{discrete} \emph{$G$-connection} on $\Gamma$ is the assignment of an element $g_e$ of the structure group $G$ to each edge of $\Gamma$. The space of connections on $\Gamma$ is therefore
\be
\mathcal{A}(\Gamma,G) \,:=\, \bigl\{ A = (g_e)_{e\in\Gamma_1}\,\in G^E\bigr\},
\ee
The curvature of a connection $A$ is encoded in its holonomies. In this discrete setting, it is thus defined as the family of $F$ group elements given by
\be \label{curvature map}
H(A)\,:=\, \Bigl( H_f(A)=\prod_{e\in\partial f} g_e^{\epsilon_{fe}} \Bigr)_{f\in\Gamma_2},
\ee
where $\epsilon_{fe}=\pm1$ is the relative orientation of the face $f$ and the edge $e$. This provides a notion of flatness on the foam: the connection is \emph{flat} if\footnote{In this paper, $\unit$ denotes the unit element of the relevant group.}
\be
H(A)\,=\,\unit.
\ee
Moreover, if $v$ and $w$ are two vertices of $\Gamma$, and $\gamma=(e_1^{\eta_1},\dots,e_n^{\eta_n})$ an edge-path connecting them, with $\eta_i=\pm1$ to take into account the orientations of the edges, we can define the \emph{parallel transport operator} from $v$ to $w$ by
\be
P_A(v,w;\gamma):=\prod_{i=1}^ng_{e_i}^{\eta_i}.
\ee
If $A$ is flat, two paths in the same homotopy class yield the same operator, and we can write simply $P_A(v,w;[\gamma])$.

A \emph{discrete gauge transformation} $h$ is a set of $V$ group elements $(h_v)_{v\in\Gamma_0}$ acting at the vertices of $\Gamma$ and mapping $A$ to
\be
\gamma_A(h) := \bigl(h_{t(e)}\,g_e\,h^{-1}_{s(e)}\bigr)_{e\in\Gamma_1}.
\ee
We denote $\mathcal{O}(\Gamma,G)_A$ the orbit of a discrete connection $A$, and $\zeta(\Gamma,G)_A:=\ker\gamma_A$ its stabilizer. When $\zeta(\Gamma,G)_A$ consists only in the center of $G^V$, we say that $A$ is \emph{irreducible}; else we say that it is \emph{reducible}.

It is important to note that the set of flat connections
\be
\mathcal{F}(\Gamma,G):=H^{-1}(\unit),
\ee
is not a manifold in general. Rather, if $G$ is algebraic, it has the structure of a real algebraic set , whose Zariski tangent space $T_\phi\mathcal{F}(\Gamma,G)$ satisfies
\be
\dim T_\phi\mathcal{F}(\Gamma,G)\leq\dim\ker dH_\phi.
\ee
The subset of $\mathcal{F}(\Gamma,G)$ saturating this inequality, however, \emph{is} a manifold, called its principal stratum. All our considerations to come are based on the assumptions that, for what concerns powercounting, its complement (`singular connections') can be neglected. We do not fully understand the generality of this assumption; see however \cite{Bonzom:2010uq} for more comments on this issue (and a counter-example).

By modding out the gauge orbits in $\mathcal{F}(\Gamma,G)$, we obtain the \emph{moduli space} of flat connections
\be
\mathcal{M}(\Gamma,G):=\mathcal{F}(\Gamma,G)/G^V.
\ee
Like $\mathcal{F}(\Gamma,G)$, the moduli space is a real algebraic set. As we will recall in sec. \ref{more}, it is a topological invariant of $\Gamma$, determined by its fundamental group.

\subsection{Twisted cohomology}\label{cohom}

If $M$ is a smooth $d$-manifold, it is well known that the exterior derivative $d^i:\Omega^i(M)\rightarrow\Omega^{i+1}(M)$ defines a cochain complex
\be
0\longrightarrow\Omega^0(M)\xrightarrow{d^0}\Omega^1(M)\xrightarrow{d^1}\dots\xrightarrow{d^{n-1}}\Omega^n(M)\xrightarrow{d^n}0,
\ee
whose cohomology $H_\textrm{dR}^*(M)$ is called the de Rham cohomology of $M$. Physically, the first de Rham cohomology group $H^1_\textrm{dR}(M)$ is the space of Maxwell fields up to gauge; more generally,  $H^i_\textrm{dR}(M)$ is the solution space of `$i$-form electrodynamics'.

There is a natural non-Abelian generalization of de Rham cohomology. If $\phi$ is flat connection on a principal $G$-bundle $P\rightarrow M$, this construction extends to forms over $M$ valued in the adjoint bundle $\Ad(P)=P\times_{\Ad}\alg$, by means of the covariant exterior derivative $d^i_ \phi:\Omega^i\big(M,\Ad(P)\big)\rightarrow\Omega^{i+1}\big(M,\Ad(P)\big)$. Indeed, the flatness of $\phi $ entails $d_ \phi ^{i+1}d_\phi^{i}=[F(A),\ \cdot\ ]=0$, which means that
\be
0\longrightarrow\Omega^0\big(M,\Ad(P)\big)\xrightarrow{d^0_\phi}\Omega^1\big(M,\Ad(P)\big)\xrightarrow{d^1_\phi}\dots\xrightarrow{d^{n-1}_\phi}\Omega^n\big(M,\Ad(P)\big)\xrightarrow{d^n}0,
\ee
is a cochain complex. The corresponding cohomology $H^*_ \phi\big(M,\Ad(P)\big)$ is the \emph{twisted} de Rham cohomology of $M$, and describes non-Abelian gauge theory in a background field $\phi$. For instance, $H_ \phi ^{d-2}(M)$ is the space of solutions  up to gauge of the equation of motion $d_ \phi ^{d-2}B=0$ of BF theory.

In the setting of cell complexes, a parallel construction is provided by \emph{cellular} cohomology. Let $\Delta$ be a finite CW complex, $C^i(\Delta)$ the free vector space over the set of $i$-cells $e^i_\alpha$, and $\delta^i:C^i(\Delta)\rightarrow C^{i+1}(\Delta)$ the cellular coboundary operator defined by
\be
\delta^i(e^i_\alpha):=\sum_{\beta}[e_\beta^{i+1},e_\alpha^i]\ e^{i+1}_\beta,
\ee
where $[e_\beta^{i+1},e_\alpha^i]$ is the incidence number of $e_\beta^{i+1}$ on $e_\alpha^{i}$. It is easy to check that $\delta^{i+1}\delta^i=0$. The \emph{cellular} cohomology groups of $\Delta$ are then $H^i(\Delta):=\ker\delta^i/\im\delta^{i-1}$. See Appendix A for a more detailed review of cellular (co)homology.

Just like de Rham cohomology, cellular cohomology can be twisted by flat $G$-connections. Let $C^i(\Delta,\alg)=C^i(\Delta)\otimes\alg\simeq\alg^{c_i(\Delta)}$ denote the set of linear combinations of $i$-cells with coefficients in the Lie algebra $\alg$, the discrete counterpart of $\Omega^i\big(M,\Ad(P)\big)$. To describe the corresponding analogue of the covariant exterior derivative, it is necessary to pick a reference vertex $v^i_\alpha$ on the boundary of each cell $e^i_\alpha$ of $\Delta$. Indeed, given two cells $e^i_\alpha$ and $e^j_\beta$ and a flat discrete connection $\phi$, we can then consider a parallel transport operator $P_\phi(v^i_\alpha,v^{i+1}_\beta)$ along a path of edges connecting $v^i_\alpha$ and $v^{i+1}_\beta$ lying on the boundary of $e^{i+1}_\beta$. (Since the connection is flat, the operator is independent of the chosen path on the boundary of $e^{i+1}_\beta$.) Via the adjoint representation of $G$, these operators act on the cochain spaces $C^i(\Delta,\alg)$, and allow to define the twisted coboundary operators $\delta_\phi^i:C^i(\Delta,\alg)\rightarrow C^{i+1}(\Delta,\alg)$ by

\beq
\delta_\phi^i(e^i_\alpha\otimes X) := \sum_{\beta}[e_\beta^{i+1},e_\alpha^i]\ \Big(e^{i+1}_\beta\otimes\Ad_{P_\phi(v^i_\alpha,v^{i+1}_\beta)}(X)\Big),
\ee
for $X\in\alg$. It can be checked that $\delta_\phi^{i+1}\delta_\phi^i=0$, and that the \emph{twisted} cellular cohomology groups $H_\phi(\Delta,\alg):=\ker\delta_\phi^i/\im\delta_\phi^{i-1}$ are well-defined, i.e. do not depend on the choice of reference vertices. Like the usual cohomology groups, they are homotopy invariants of $\Delta$; their dimensions $b^i_\phi(\Delta,\alg):=\dim H_\phi(\Delta,\alg)$ are the twisted Betti numbers of $\Delta$. Let us emphasize that, in general, these are not simply related to the standard Betti numbers $b^i(\Delta)$. There \emph{is} such a simple relationship, however, for the twisted Euler characteristic, which satisfies
\be\label{characteristic}
\chi_\phi(\Delta,\alg):=\sum_{i=0}^d(-1)^i\dim C^i(\Delta,\alg)=(\dim G)\sum_{i=0}^d(-1)^ic_i(\Delta)=(\dim G) \chi(\Delta).
\ee
Here $c_i(\Delta)$ is the number of $i$-cells of $\Delta$. By the Euler-Poincar\'e theorem, it follows that
\be
\sum_{i=0}^d(-1)^ib^i_\phi(\Delta,\alg)=(\dim G) \chi(\Delta).
\ee
This property will prove useful to analyze the divergence degree of a complex in topological terms (next section).

The relevance of the twisted cellular cohomology lies in its natural gauge-theoretic interpretation, which is strictly analogous to the continuous case. Indeed, a moment of reflection shows that $\delta^0_\phi$ is the differential of the gauge transformation map $\gamma_\phi$ at the identity,
\be
\delta^0_\phi=(d\gamma_\phi)_1.
\ee
Thus, the space of $1$-coboundaries $ \im \delta^0_\phi$ is the tangent space of the gauge orbit of $\phi$, and the space of $0$-cocycles $\ker\delta^0_\phi$ is the Lie algebra of its stabilizer $\zeta(\Gamma,G)_\phi$. In particular, $H^0_\phi$ vanishes if and only if $\phi$ is irreducible. Similarly, one sees that
\be
\delta^1_\phi=dH_\phi.
\ee
Hence, the space of $1$-cocycles $\ker\delta_\phi^1$ is the tangent space to $\mathcal{F}(\Gamma,G)$ at $\phi$, and the first cohomology group $H_\phi^1(\Delta,\alg)$ is the tangent space to $\mathcal{M}(\Gamma,G)$ at $\mathcal{O}(\Gamma,G)_\phi$.

The reader might wonder whether the higher coboundary operators also have a gauge-theoretic interpretation. From the perspective of discrete connections on $\Delta$, the answer is negative. However, the flat spinfoam model can be analyzed as a discrete BF model, with an additional variable in $\alg$ on each face of $\Delta$; the higher coboundary operators turn out to describe the (reducible) shift gauge symmetry of the discrete $B$ field. We will detail this point of view in \cite{discreteBF}.

\subsection{Divergence degree from Laplace approximation}

In our previous papers \cite{Bonzom:2010fk,Bonzom:2010uq}, we introduced a heat kernel regularization for the partition function \eqref{partition},
\be\label{regularized}
\calZ_\tau(\Gamma,G) := \int_{\cal A}dA\ \prod_{f\in\Gamma_2} \k\bigl(H_f(A)\bigr),
\ee
where $dA$ is the Haar measure on $G^E$. Each delta function of \eqref{partition} is replaced by the heat kernel $\k$ on $G$. As is apparent from the Peter-Weyl expansion of $\k$,
\be
\k(g)=\sum_{j=0}^\infty d_je^{-4\tau j(j+1)}\chi_j(g),
\ee
where $d_j$ is the dimension of the irreducible representation $j$, and $\chi_j$ its character, this regularization amounts to cutting the range of spins off to $\Lambda_\tau\propto\tau^{-1/2}$. This is analogous to the more commonly used `sharp' regularization of the group delta function $\delta_\Lambda(g)=\sum_{j=0}^\Lambda d_j\chi_j(g)$, but has the advantage over the latter of being positive. This feature allows to write the heat kernel $\k$ as a real exponential, a form which is suitable for a Laplace approximation of the integral \eqref{regularized}. Indeed, we have the `small time' heat kernel asymptotics \be\k(g) \u{\tau\rightarrow 0}{\sim} (4\pi\tau)^{-\f{\dim G}{2}}\ e^{-\f{|g|^2}{4\tau}},\ee where $\lvert g\rvert$ is the Riemannian distance in $G$ from the identity to $g$. This formula motivates the definition of \emph{divergence degree} of $\Delta$ as the number $\Omega(\Delta,\alg)$ such that the limit
\be
\calZ'(\Gamma,G):=\u{\tau\rightarrow 0}{\lim}\ (4\pi\tau)^{\Omega(\Gamma,G)/2} \calZ_{\tau}(\Gamma,G)
\ee
is finite and non-vanishing. Thus, we can write \eqref{regularized} as a Laplace integral,
\be
\calZ_\tau(\Gamma,G)\u{\tau\rightarrow 0}{\sim}(4\pi\tau)^{-\f{(\dim G)F}{2}}\int_{\mathcal{A}(\Gamma,G)}dA\ \exp\Biggl(-\sum_{f\in\Gamma_2}\f{\lvert H_f(A)\rvert^2}{4\tau}\Biggr).
\ee
Following Forman, we introduced Riemannian normal coordinates in a tubular neighborhood of $\mathcal{F}(\Gamma,G)$, on which this integral is supported, and used them to split the integral into the tangential and normal directions to $\mathcal{F}(\Gamma,G)$. The normal integrals are Gaussians with variance $\tau$, and the tangential ones do not depend on $\tau$. Hence, the divergence degree is readily identified as
\be
\Omega(\Gamma,G)=(\dim G)F-\textrm{codim}\ \mathcal{F}(\Gamma,G),
\ee
in which $\textrm{codim}\ \mathcal{F}(\Gamma,G):=\dim\mathcal{A}(\Gamma,G)-\dim\mathcal{F}(\Gamma,G)$ is the number of directions normal to $\mathcal{F}(\Gamma,G)$.\footnote{If the algebraic set $\mathcal{F}(\Gamma,G)$ is not irreducible, the divergence degree might take different values on its different irreducible components.} But since the tangent space to $\mathcal{F}(\Gamma,G)$ can be described as the space of $1$-cocycles in the twisted cohomology, we have $\textrm{codim}\ \mathcal{F}(\Gamma,G)=\rk\delta_\phi^1$ for any non-singular connection $\phi$, and therefore
\be
\Omega(\Gamma,G)=(\dim G)F-\rk\delta_\phi^1=b^2_\phi(\Gamma,G),
\ee
That is, the divergence degree of the complex $\Gamma $ is given by its \emph{second twisted Betti number}. When $\Gamma $ is simply connected, or when $G$ is Abelian, this formula reduces to \cite{Bonzom:2010fk}
\be
\Omega(\Gamma,G)=(\dim G)b^2(\Gamma),
\ee
in which $b^2(\Gamma)$ is the cellular second Betti number.

\subsection{Sorting out topology from cell structure}

Part of the interest of this result is that it holds for arbitrary $2$-complexes. If, however, we assume that $\Gamma$ is the $2$-skeleton of the cell decomposition $\Delta_M$ an $n$-pseudomanifold $M$, this result can be sharpened considerably. Consider indeed the twisted Euler-Poincar\'e formula for the $2$-skeleton $\Gamma$,
\be
\chi_\phi(\Gamma,G)=b_\phi^0-b_\phi^1+b_\phi^2.
\ee
From the discussion at the end of sec. \ref{cohom}, we know that $b^0_\phi$ is the dimension of the common stabilizer $\zeta(\Gamma,G)$ of non-singular connections, and that $b_\phi^1$ is the dimension of the moduli space $\mathcal{M}$. Combining this with \eqref{characteristic}, we get
\be
\Omega(\Gamma,G)=\dim\mathcal{M}(\Gamma,G)-\dim\zeta(\Gamma,G)+(\dim G)\chi(\Gamma).
\ee
Moreover, by definition of the Euler characteristic, we have
\be
\chi(\Gamma)=V-E+F=\chi(\Delta_M)-\sum_{i=3}^n(-1)^ic_i(\Delta_M).
\ee
Hence,
\be\label{resultat}
\Omega(\Gamma,G)=\dim\mathcal{M}(\Gamma,G)-\dim\zeta(\Gamma,G)+(\dim G)\chi(\Delta_M)-(\dim G)\sum_{i=3}^n(-1)^ic_i(\Delta_M).
\ee
Thus, we have a decomposition
\be\label{decomp}
\Omega(\Gamma,G)=I(M,G)+\omega(\Delta_M,G),
\ee
where
\be\label{topological}
I(M,G):=\dim\mathcal{M}(\Gamma,G)-\dim\zeta(\Gamma,G)+(\dim G)\chi(M).
\ee
is the `topological' part of the divergence degree, depending only on the fundamental group of $M$, and
\be\label{cellular}
\omega(\Delta_M,G):=-(\dim G)\sum_{i=3}^n(-1)^ic_i(\Delta_M)
\ee
is its `cellular' part, depending on the given cell decomposition $\Delta_M$ of $M$.\footnote{Note that this decomposition actually holds for any $n$-dimensional cell complex. The restriction to pseudo-manifolds is made for comparison with the literature and as a natural choice in group field theory.} This decomposition is what we mean by ``sorting out topology from cell structure".

\subsection{More on the evaluation of $\Omega(\Gamma,G)$}\label{more}
An important feature of the above decomposition of the divergence degree is its computational propitiousness. Indeed,
\begin{itemize}
\item the cellular part $\omega(\Delta, G)$ can be evaluated straightforwardly by counting the cells of the complex, without solving any equation,
\item the topological part $I(M,G)$ is determined $(i)$ by the Euler characteristic, and $(ii)$ by the fundamental group $\pi_1(M)$, as we will now explain. As such, it can be evaluated using any presentation of $\pi_1(M)$.
\end{itemize}


Since the moduli space of flat connections is by definition $\mathcal{M}(\Gamma,G) = \mathcal{F}(\Gamma,G)/G^V$, we wish to factor out the gauge transformations, i.e. to identify gauge equivalent flat connections. To this effect, it is convenient to reduce gauge transformations so that they act at \emph{one} vertex only. This can be done using the standard process of contracting a spanning tree of $\Gamma$. This process does not affect the amplitude \eqref{partition} since the integrand is invariant under gauge transformations.

We can thus formulate the partition function on a deformation retract of $\Gamma$ along a spanning tree $T$, denoted $\Gamma/T$, with a single vertex, $E-V+1$ edges, and the same number of faces (and higher-dimensional cells in $\Delta_M$). This new complex carries the residual part of the gauge transformations: the conjugation of the elements $g_e$ by a single element $h\in G$,
\be
h\,\cdot\,A = \bigl(h\, g_e\, h\mone\bigr)_{e\in(\Gamma/T)_1}.
\ee

This reduction is a useful trick to describe the moduli space $\mathcal{M}(\Gamma,G)$ of flat connections on $\Gamma$ in terms of its fundamental group $\pi_1(\Gamma) = \pi_1(\Gamma/T)$. Indeed, the retraction of $T$ in $\Gamma$ naturally provides a presentation of its fundamental group: the generators $a_{e}$ correspond to the remaining edges $e$ in $\Gamma/T$, and there is one relation per face, expressing the contractibility of paths along faces, similarly to the flatness condition on the discrete connection:
\be\label{standardpresentation}
\pi_{1}(\Gamma)=\bra(a_{e})_{e\in(\Gamma/T)_{1}}\
\vert\ (\prod_{e}a_{e}^{\epsilon_{fe}})_{f\in(\Gamma/T)_{2}}=\unit \ket.
\ee
(Notice that the relationship between presentations of groups and 2-complexes goes both ways: a finite presentation of a group $\pi$ unambiguously determines a 2-complex $\Gamma$. From a single vertex, draw an edge for each generator, and attach the faces according to the relators.\footnote{Note that trivial relations such as $aa^{-1}=\unit$ must not be eliminated from the presentation of $\pi$ for this duality to hold. An example of this issue is provided by the `dunce hat': while $\bra a\vert a^2a^{-1}=\unit\ket$ is obviously equivalent to $\bra a\vert a=\unit\ket$ as a group presentation, the corresponding $2$-complexes, the dunce hat and the disc respectively, are not.})

This argument shows that a flat connection on $\Gamma/T$ can be seen as a homomorphism from $\pi_1(\Gamma)$ to $G$,
\be \label{representation space}
\mathcal{F}(\Gamma/T,G)\simeq \Hom\bigl(\pi_1(\Gamma),G\bigr).
\ee
This space is usually called the \emph{representation variety} of $\pi_1(\Gamma)$ into $G$ (or the space of flat $G$-bundles over $M$, when $\pi_1(\Gamma)$ is seen as the fundamental group of a manifold $M$). Moreover, the moduli space of flat connections is the variety of characters, i.e. the quotient of this representation variety by the remaining group action on $\Gamma/T$ (conjugation),
\beq
\mathcal{M}(\Gamma,G) = \Hom\bigl(\pi_1(\Gamma),G\bigr)/G.
\ee
Obviously, the moduli space of flat connections of $\Gamma/T$ is the same as that of $\Gamma$.

Now, recall that to compute $I(M, G)$, we need two quantities in addition to the Euler charateristic: $\dim \mathcal{M}(\Gamma,G)$ and $\dim\zeta(\Gamma,G)$. Both can be evaluated from any presentation of the fundamental group, and in particular from \eqref{standardpresentation}. Often this presentation can be simplified by formal manipulations, or by identifying the isomorphism class of the group. Since the dimension of the moduli space is 
\beq \label{dimM}
\dim\mathcal{M}(\Gamma,G) = \dim \Hom\bigl(\pi_1(\Gamma),G\bigr)- \dim\mathcal{O}(\Gamma,G),
\ee
where $\mathcal{O}(\Gamma,G)$ is an orbit under the $G$-action, the orbits should also be described. The last ingredient is the dimension of the stabilizer of a representation of $\pi_1(\Gamma)$. This is simply
\beq \label{dimzeta}
\dim\zeta(\Gamma,G) = \dim G - \dim \calO.
\ee
Note that it may be that different non-singular flat connections on $\Gamma/T$ lead to different values of $I$. These correspond to different irreducible components of the variety of characters. For small time $\tau$, the leading behaviour of the partition function $\calZ_\tau$ is obviously given by the largest value of $I$.

\section{Comparison with other results}\label{comparison}

The previous powercounting results for the flat spinfoam model dealt with the ``simplicial" Boulatov-Ooguri complexes exclusively.\footnote{This means that, in dimension $d$, exactly $(d+1)$ edges (resp. $d$ faces) are incident on each vertex (resp. edge) of $\Gamma$, corresponding to the number of faces of a $d$-simplex (resp. $(d-1)$-simplex).} They fall in four classes:

\medskip\begin{itemize}
\item
\emph{Powercounting from Pachner moves}. If one assumes that $\Gamma$ arises from a triangulation $T_M$ of a closed $n$-manifold $M$, as the $2$-skeleton of its dual cell complex, one can see that Pachner moves on $T_M$ generate divergences in the partion function. Ponzano and Regge \cite{PR}, Boulatov \cite{Boulatov:1992vp} and Ooguri \cite{Ooguri:1992eb} relied on this observation to relate the divergence degree to the combinatorics of $\Gamma$.
\item
\emph{Powercounting from bubble counting}. A different approach to the divergences of the flat spinfoam model was initiated by Perez and Rovelli \cite{Perez:2000fs}, who realized that they are related to the presence of ``bubbles" within $\Gamma$. Freidel and Louapre \cite{Freidel:2004vi,Freidel:2004nb}, and later Freidel, Gurau and Oriti \cite{Freidel:2009hd}, pushed this intuition further in three-dimensions and obtained a powercounting estimate for certain special complexes, coined ``type 1". Within Gurau's colored tensor models, this result was then extended to higher dimensions by Ben Geloun \emph{et al.} \cite{Geloun:2010nw}.

\item
\emph{Powercounting from vertex counting} \cite{Magnen:2009at,Geloun:2009pe}. Yet another approach relies on the field-theoretic notion that the divergence degree of a Feynman diagram can be bounded by the number of its vertices. Such bounds were obtained by Magnen \emph{et al.} in the Boulatov model \cite{Magnen:2009at}, and adapted to the colored tensor models in \cite{Geloun:2009pe}.

\item
\emph{Powercounting from jackets}. The notion of ``jacket" for a (colored) Boulatov-Ooguri complex was introduced by Ben Geloun \emph{et al.} in \cite{Geloun:2010nw}, and used by Gurau and Rivasseau \cite{Gurau:2011aq} to obtain an upper bound on the divergence degree improving the one obtained by vertex counting mentioned above.
\end{itemize}

Before we go on and show how our main formula \eqref{decomp} encompasses all the results quoted above, we would like to stress an important point: except for the perturbative bounds obtained from vertex counting, these results are \emph{all} based on the (usually implicit) assumption that $\Omega(\Gamma,G)$ is an integer multiple of $\dim G$. An immediate consequence of \eqref{topological} and \eqref{cellular}, however, is that this is not true in general, because neither $\dim\mathcal{M}(\Gamma,G)$ nor $\dim\zeta(\Gamma,G)$ in the topological part $I(M,G)$ are multiples of $\dim G$.

\subsection{Powercounting from Pachner moves}

\subsubsection{Ponzano-Regge scaling in three dimensions}

When Ponzano and Regge introduced their model in 1968, they considered a $3$-manifold $M$ equipped with a triangulation $T_M$, $G=\SU(2)$, and conjectured that the divergence degree was given by three (i.e. $\dim \SU(2)$) times the number of vertices $V(T)$ in the triangulation. They were guided in making this conjecture by the following consideration: because of the Biedenharn-Elliot identity for $6j$-symbols, the formal partition functions of two triangulations $T_M$ and $P_{14}T_M$ of $M$ related by a $1-4$ Pachner move satisfy
\be
\calZ\big(P_{14}T_M\big)=\Big(\sum_{j=0}^{\infty}(2j+1)^2\Big)\calZ\big(T_M\big).
\ee
Since $\sum_{j=0}^{\Lambda}(2j+1)^2$ scales as $\Lambda^3$ when $\Lambda\rightarrow\infty$, and since the $1-4$ move introduces a single new vertex in the triangulation, the conclusion that each vertex contributes a factor of $\Lambda^3$ in the partition function appears tentalizing. Indeed, Ponzano and Regge proposed to cut off the sums in their state-sum to a maximal value $\Lambda$ to obtain a finite value $\calZ_\Lambda\big(T\big)$ for the partition function, and tentatively defined a regularized partition function by
\be
\mathcal{Z}'_{\textrm{PR}}(T_M)=\underset{\Lambda\rightarrow\infty}{\lim}\ \Lambda^{-3V(T_M)}\calZ_\Lambda\big(T_M\big).
\ee

The origin of the intuition that divergences in the Ponzano-Regge model are related to the vertices of $T_M$, but also the reason why this naive regularization is bound to fail, is completely elucidated by our decomposition \eqref{decomp}. Indeed, taking $\Delta_M$ as the dual cell complex to a triangulation $T_M$, and noting that in three dimensions, a vertex is dual to a $3$-cell, we see that
\be
\omega(T_M)=3c_3(\Delta_M)=3V(T_M).
\ee
That is, the relationship between divergences of $\calZ(T_M)$ and vertices of $T_M$ conjectured by Ponzano and Regge is exact for the non-topological part of $\Omega(T_M)$, but misses completely its topological part $I(M)$. This is because the argument based on the $1-4$ Pachner move actually estimates not the divergence degree itself, but its \emph{variation} in the move, that is
\be
\Omega(P_{14}T_M)-\Omega(T_M)=\omega(P_{14}T_M)-\omega(T_M)=3\big(V(P_{14}T_M)-V(T_M)\big),
\ee
in which $I(M)$ \emph{cancels}.


\subsubsection{Ooguri scaling in four dimensions}

A similar observation was made in the four-dimensional case by Ooguri in \cite{Ooguri:1992eb}. He considered separately the $1-5$ move (in which the number of vertices and edges increase by one and five respectively) and the $3-3$ move (in which the number of vertices is left unchanged while the number of edge is increased by one), and found that $\mathcal{Z}(T_M)$ depends on the triangulation $T_M$ only through a divergent factor measured by $3\big(E(T_M)-V(T_M)\big)$, where $V(T_M)$ and $E(T_M)$ are the number of vertices and edges of $T_M$. Although he could not show this result rigorously (because he could not make sense of the partition function as a finite number), we can interpret his argument along the same lines as in three dimensions: it provides the correct non-topological divergent degree $\omega(T_M)$, but misses the topological part $I(M)$. Indeed, in the dual complex $\Delta_M$ to $T_M$, there is one $4$-cell per vertex, and one $3$-cell per edge in $T_M$, so that
\be
\omega(\Delta_M)=3\big(c_3(\Delta_M)-c_4(\Delta_M)\big)=3\big(E(T_M)-V(T_M)\big).
\ee

\subsection{Powercounting from bubble counting}

\subsubsection{Freidel-Gurau-Oriti type $1$ graphs in three dimensions}

Freidel, Gurau and Oriti defined in \cite{Freidel:2009hd} a restricted class of $2$-complexes $\Gamma$, which they called ``type $1$", for which they could show that the divergence degree is simply related to the number $B(\Gamma)$ of ``bubbles" in $\Gamma$,
\be\label{FGObubbles}
\Omega(\Gamma)=3\big(B(\Gamma)-1\big).
\ee
With the view that $\Gamma$ is the $2$-skeleton of a cell complex $\Delta_M$ decomposing a $3$-pseudomanifold $M$, the ``bubbles" of $\Gamma$ in the sense of Freidel \emph{et al.} correspond to the $3$-cells of $\Delta_M$, hence \eqref{FGObubbles} reads
\be\label{FGO}
\Omega(\Delta_M)=3(c_3(\Delta_M)-1).
\ee
Taking their cues from the Turaev-Viro model, they speculated that the ``type $1$" condition should correspond to ``topologically trivial manifolds", i.e. to the $3$-sphere. Given the decomposition \eqref{decomp}, the identity \eqref{FGO} is equivalent to
\be
I(M)=\dim\mathcal{M}(\Gamma,G)-\dim\zeta(\Gamma,G)=-3.
\ee
In other words, the ``type 1" complexes are, for a given number of $3$-cells in the corresponding complex $\Delta_M$, are \emph{minimally} divergent. Moreover, they are such that their flat connections up to gauge are isolated and completely reducible. We will see in the next section that this last condition alone does not single out the $3$-sphere (it is also satisfied \emph{e.g.} by the real projective space $\mathbb{R}P^3$).

\subsubsection{Abelian powercounting in the colored model}

In \cite{Geloun:2010nw}, Ben Geloun \emph{et al.} considered Gurau's \emph{colored model}, and used the corresponding ``bubble homology" to obtain the following formula for the divergence degree in dimension $d\geq3$ for the \emph{Abelian} structure groups $G=\mathbb{R}$:\footnote{That $\mathbb{R}$ is non-compact introduces another source of divergences, which can be tamed with a second cutoff.}

\be\label{linearized}
\Omega(\mathcal{G})=\sum_{k=3}^{d+1}(-1)^{k-1}c_k(\mathcal{G})+\sum_{k=2}^{d-1}(-1)^kb_k(\mathcal{G}).
\ee
Here, $\mathcal{G}$ is a $(d+1)$-colored graph, $c_k(\mathcal{G})$ is for $0\leq n\leq d$ the number of $k$-bubbles in $\mathcal{G}$ , $c_{d+1}(\mathcal{G}):=1$ and $b_k(\mathcal{G})$ is the $k$-th ``bubble" Betti number of $\mathcal{G}$. All the details are given in Appendix \ref{colored}.

The relationship with our result is based on the following correspondence. A $(d+1)$-colored graph $\mathcal{G}$ naturally defines a $d$-dimensional CW complex $\Delta_{\mathcal{G}}$ with the following property:
\medskip\begin{itemize}
\item
the underlying graph of $\mathcal{G}$ is the $1$-skeleton of $\Delta_{\mathcal{G}}$,
\item
the number of $k$-bubbles of $\mathcal{G}$ equals the number of $k$-cells of $\Delta_{\mathcal{G}}$,
\item
the ``bubble" homology of $\mathcal{G}$ coincides with the cellular homology of $\Delta_{\mathcal{G}}$.
\end{itemize}
With this correspondence, the relationship between \eqref{linearized} and our result \eqref{resultat} is straightforward. First, note that what is denoted $c_{d+1}(\mathcal{G})$ in \eqref{linearized} is really $b_d(\mathcal{G})=1$, so that it can (and should) be rewritten

\be
\Omega(\mathcal{G})=\sum_{k=3}^{d}(-1)^{k-1}c_k(\mathcal{G})+\sum_{k=2}^{d}(-1)^kb_k(\mathcal{G}).
\ee
Then, use the identities $c_k(\mathcal{G})=c_k(\Delta_{\mathcal{G}})$ and $b_k(\mathcal{G})=b_k(\Delta_{\mathcal{G}})$ and the Euler-Poincar\'e relation
\be
\chi(\Delta_{\mathcal{G}})=\sum_{k=0}^d(-1)^kb_k(\Delta_{\mathcal{G}})
\ee
to get
\be
\Omega(\mathcal{G})=\sum_{k=3}^{d}(-1)^{k-1}c_k(\Delta_{\mathcal{G}})+\chi(\Delta_{\mathcal{G}})-b_0(\Delta_{\mathcal{G}})+b_1(\Delta_{\mathcal{G}}).
\ee
This is the same as \eqref{resultat} in this particular case. Our result can therefore be described as the generalization of \eqref{linearized} to general cell complexes, and non-Abelian groups.

\subsection{Powercounting from vertex and jacket counting} \label{jacket-counting}

Using tools from perturbative quantum field theory, and notably Cauchy-Schwarz inequalities, an upper bound on the divergence degree in the Boulatov model was obtained in \cite{Magnen:2009at}. It is formulated in terms of the number of vertices rather than $3$-cells in the complex, and reads for a Boulatov complex $\Gamma$ without generalized tadpole \cite{Geloun:2009pe}
\be
\Omega(\Gamma)\leq \f{3}{2}c_0(\Gamma)+6.
\ee
Within the colored model, this bound was then generalized to higher dimensions in \cite{Geloun:2009pe}. In dimension $d$, it becomes for a colored graph $\mathcal{G}$ with $V(\mathcal{G})$ vertices
\be\label{bound}
\Omega(\mathcal{G})\leq \f{3(d-1)(d-2)}{4}V(\mathcal{G})+3(d-1)
\ee
Using the notion of ``jacket" introduced in \cite{Geloun:2010nw}, Gurau and Rivasseau further improved this bound in \cite{Gurau:2010ba,Gurau:2011aq}, obtaining
\be\label{jacketbound}
\Omega(\mathcal{G})\leq \f{3(d-1)(d-2)}{4}V(\mathcal{G})+3(d-1)-\f{6(d-2)}{d!}\sum_{\mathcal{J}}g(\mathcal{J}),
\ee
where $g(\mathcal{J})$ is the genus of (orientable surface dual to) the jacket $\mathcal{J}$.

Remarkably, the jacket bound \eqref{jacketbound} follows easily from our exact result \eqref{resultat}. Indeed, denoting $E(\mathcal{G})$ the number of edges and $F(\mathcal{G})$ the number of faces of $\mathcal{G}$, the formula \eqref{resultat} reads
\be
\Omega(\mathcal{G})=\dim\mathcal{M}(\Gamma,G)-\dim\zeta(\Gamma,G)+3\big(V(\mathcal{G})-E(\mathcal{G})+F(\mathcal{G})\big).
\ee
Since a colored graph has no tadpole, we have $2E(\mathcal{G})=(d+1)V(\mathcal{G})$, hence
\be\label{intermediate}
\Omega(\mathcal{G})\leq\dim\mathcal{M}(\Gamma,G)-\f{3(d-1)}{2}V(\mathcal{G})+3F(\mathcal{G}).
\ee
Moreover, since
\be
\dim\mathcal{M}(\Gamma,G)\leq\dim\textrm{Hom}\big(\pi_1(\mathcal{G}),\SU(2)\big),
\ee
and $\pi_1(\mathcal{G})$ is a subgroup of $\pi_1(\mathcal{J})$ for each jacket $\mathcal{J}$ of $\mathcal{G}$, we have
\be
\dim\mathcal{M}(\Gamma,G)\leq\f{1}{J(\mathcal{G})}\sum_{\mathcal{J}}\dim\textrm{Hom}\big(\pi_1(\mathcal{J}),\SU(2)\big),
\ee
where
\be
J(\mathcal{G})=\f{d!}{2}
\ee
is the number of jackets of $\mathcal{G}$. Now it is well-known that the dimension of the $\SU(2)$ representation variety of a genus $g$ surface is $6g-3$. Thus,
\be\label{moduligenus}
\dim\mathcal{M}(\Gamma,G)\leq\f{2}{d!}\sum_{\mathcal{J}}\big(6g(\mathcal{J})-3\big).
\ee
Using the Euler relation
\be
\chi(\mathcal{J})=2-2g(\mathcal{J})=V(\mathcal{J})-E(\mathcal{J})+F(\mathcal{J}),
\ee
the combinatorial facts that $V(\mathcal{J})-E(\mathcal{J})=\f{d-1}{2}V(\mathcal{G})$ and that each face of $\mathcal{G}$ belongs to $(d-1)!$ jackets, we have \cite{Gurau:2011xq}
\be
F(\mathcal{G})=\f{d(d-1)}{4}V(\mathcal{G})+d-\f{2}{(d-1)!}\sum_{\mathcal{J}}g(\mathcal{J}).
\ee
Using this relation in \eqref{moduligenus} gives a bound on $\dim\mathcal{M}(\Gamma,G)$ which, when inserted in \eqref{intermediate}, immediately gives the jacket bound \eqref{jacketbound}.

\section{Three-dimensional examples}\label{examples}

In this section, we illustrate how the divergence degree, and more specifically its topological part $I(M,G)$, can be computed in certain three-dimensional cases, with $G=\SU(2)$. (Hence, we will drop the reference to $G$ in the notation.) Some of them cannot be handled by the previous methods and do not saturate the corresponding bounds \cite{Freidel:2004vi,Barrett:2009ys,Freidel:2009hd,Geloun:2010nw}.

Three-dimensional closed manifolds have $\chi(M)=0$, so $I(M)=\dim\mathcal{M}-\dim\zeta$ and it is completely determined by the fundamental group. It is also well-known that $\calM$ is a finite set of points, and hence $\dim\calM=0$, when the fundamental group $\pi_1(M)$ is finite (but we will re-derive this feature in specific examples for illustrative purpose).

The method to compute $I(M)$ for a cellular pseudomanifold (and in fact for any cell complex) has been presented in the section \ref{more}. The main steps are:
\begin{itemize}
\item Choose a presentation of $\pi_1(M)$ (e.g. from a deformation retract of the 2-skeleton $\Gamma$).
\item Identify the representation variety $\Hom\bigl(\pi_1(M),G\bigr)$ using the chosen presentation. In general, it has several irreducible components.
\item Identify the orbit of the $G$-action on each irreducible component.
\item Apply the formula \eqref{dimM} and \eqref{dimzeta}.
\end{itemize}


\subsection{The $3$-torus}

The case of the $3$-torus $T^3$ was discussed in Appendix C of \cite{Freidel:2004vi}. Its fundamental group has the presentation
\be
\pi_1(T^3)=\bra a,b,c\ \vert\ [a,b]=[a,c]=[b,c]=1\ket.
\ee
The representations of this group in $\SU(2)$ are of the form
\be
\phi=\Big(\exp(\psi_a\hat{n}.\vec{\tau}),\ \exp(\pm\psi_b\hat{n}.\vec{\tau}),\ \exp(\pm\psi_c\hat{n}.\vec{\tau})\Big)\in\SU(2)^3,
\ee
with $\hat{n}\in S^2$ their common direction of rotation, $\psi_{a,b,c}\in[0,\pi]$ three class angles, and $\vec{\tau}$ the 3-vector formed by a set of (anti-Hermitian) generators of the algebra $\su(2)$. They form a $5$-dimensional manifold $\calF$.

The group action by conjugation rotates the direction $\hat{n}$. For each representation there is a stabilizer $\zeta$ isomorphic to $\U(1)$ (and larger if the three rotations are in the center of $\SU(2)$) which leaves it invariant: the subgroup generated by the direction $\hat{n}$. Hence, the dimension of the stabilizer is $\dim \zeta=1$. Thus
\begin{eqnarray}
\dim\mathcal{M}&=&5-2=3,\nonumber\\ \dim\zeta&=&1.
\end{eqnarray}
Hence,
\be
I(T^3)=2.
\ee
Note that this implies that the divergence degree of a cell decomposition of $T^3$ cannot be a multiple of $\dim\SU(2)=3$, as the procedure of \cite{Freidel:2004vi} assumed implicitly.

\subsection{Lens} \label{sec:lens-prism}

Lens spaces $L_{p,q}$ are standard spherical manifolds, with $\Z_p$ as their fundamental group (we exclude the case of $L(0,1)= S^1\times S^2$ which can be understood separately). They include as a particular case the real projective space $\mathbbm{R P}^3$, which is the first example in the appendix of \cite{Gurau:2009tw}.

The standard presentation of $\Z_p$ is of course
\be
\Z_p=\bra a\ \vert\ a^p=1\ket.
\ee
If $0<r<\pi$, let $S^2_r$ be the subset of $\SU(2)$ defined by
\be \label{2-sphere}
S^2_r=\Bigl\{\exp\bigl(r\,\hat{n}.\vec{\tau}\bigr)\ ; \ \hat{n}\in S^2\Bigr\},
\ee
consisting in those rotations of fixed angle $r$. In the topological picture induced by the identification $\SU(2)\simeq B^3/\pp B^3$, $S^2_r$ is a sphere of radius $r$ centered on the origin. (In this picture,  the central elements $\pm\unit$ of $\SU(2)$ are the center and boundary of $B^3$ respectively.)

Hence,
\begin{itemize}
\item If $p=1$ (the $3$-sphere), there is only one representation, the trivial one.
\item If $p=2$ (the real projective space), the representations of $\Z_2$ send its generator to an element of the center of $\SU(2)$, i.e. $\{\pm \unit\}$.
\item If $p\geq 3$, the set of representations decomposes as $\Hom (\Z_p, \SU(2)) = \calF_p^{(0)}\sqcup\calF_p^{(2)}$, with
\begin{eqnarray} \calF_p^{(0)} & = & \left\{ \begin{array}{ll}
\{\unit\}  & \qquad\textrm{$p$ odd}\\ \zeta(\SU(2))=\{\pm\unit\}  & \qquad\textrm{$p$ even}\end{array} \right. \\ \calF_p^{(2)} & = & \bigsqcup_{1\leq k< p/2}S^2_{\f{2\pi k}{p}}  \end{eqnarray}
This means that the class angles admit a finite number of values, $r= 2k\pi/p$, for $k=0,\dotsc,\lfloor \f{p}2\rfloor$. The set $\calF_p^{(0)}$ consists of points, while $\calF_p^{(2)}$ is a union of $2$-spheres.
\end{itemize}

\medskip

The next step consists in finding the orbits generated by conjugation in $\SU(2)$. It turns out that all of the above representations are reducible.

\begin{itemize}
\item
In the cases $p=1,2$, and on $\calF_p^{(0)}$ when $p\geq3$, the representations are left invariant by the group action (since they commute with the whole group): they are central. Thus, the centralizer is $\zeta = \SU(2)$, which is 3-dimensional.
\item
On $\calF_p^{(2)}$ for $p\geq 3$, group conjugation corresponds to rotation of the axis $\hat{n}$. Similarly to the 3-torus case, the stabilizer is then $\zeta=\U(1)$.
\end{itemize}

In all cases, the moduli space $\calM$ consists in $p$ distinct points, one for each connected component of $\Hom(\Z_p,\SU(2))$, hence $\dim\calM=0$. However, when $p\geq 3$, one can compute two different values of the topological part of the divergence degree, one on $\calF_p^{(0)}$ and another on $\calF_p^{(2)}$. The relevant value is obviously that which gives the most divergent contribution to the partition function and thus the greatest value of $I= \dim \calM-\dim\zeta$. We get: $I = -3$ on $\calF_p^{(0)}$, compared with $I=-1$ on $\calF_p^{(2)}$. This means that the relevant value is the one computed on the less reducible representations, i.e. those with the smallest stabilizer.

In conclusion
\be
I\big(L_{p,q}\big)= \begin{cases} -3  & \text{if $p=1,2$}\\
 -1  & \text{if $p\geq3$.}\end{cases}
\ee

\subsection{Prism spaces}

Prism manifolds $P_{m,n}$, with $m\geq 1,n\geq 2$, form a different class of spherical manifolds. They are characterized by their fundamental group
\be
\pi_1(P_{m,n})=\bra x,y\ \vert\ xyx^{-1}=y\mone,\ x^{2m}=y^{n}\ket.
\ee
Let us first give the irreducible representations $\phi$ in the case of $m$ even. The first relation imposes the direction of the rotation $\phi(x)$ to be orthogonal to that of $\phi(y)$, and its class angle to be $\f{\pi}{2}$. Then the second relation reduces to $y^n=\unit$, which only constrains the class angle of $\phi(y)$ to be $\psi_y = 2k\pi/n$ for $k=1,\dotsc,\lfloor \f{n}{2}\rfloor$ (the case $k=0$ gives a reducible representation). This way the set of irreducible representations is identified as
\beq
\calF_{m,n}^{\rm irr} = \Bigl\{ \phi(x) = \exp\bigl(\f{\pi}{2} \hat{n}_x\cdot\vec{\tau}\bigr),\ \phi(y) = \exp\bigl(\f{2k\pi}{n} \hat{n}_y\cdot\vec{\tau}\bigr)\ ;\ k=1,\dotsc,\lfloor \f{n}{2}\rfloor,\ (\hat{n}_x, \hat{n}_y)\in (S^2)^2, \hat{n}_x\cdot \hat{n}_y = 0\Bigr\}
\ee
Clearly this space is of dimension 3. Moreover, simultaneous conjugation of $\phi(x), \phi(y)$ by some $\SU(2)$ element induces a simultaneous rotation on $\hat{n}_x, \hat{n}_y$. Thus, the orbit is 3-dimensional, isomorphic to $\SU(2)/\{\pm \unit\}=\SO(3)$ and the centralizer is just the center of $\SU(2)$. This leads to $\dim \calM = 0, \dim\zeta=0$, and hence:
\beq
I(P_{m,n}) = 0,
\ee
when $m$ is even, and evaluated on irreducible representations. The situation is similar for $m$ odd, only the specific values of the class angle of $\phi(y)$ are changed.

As for reducible representations, they are obtained by taking $\phi(y)=\pm\unit\in\zeta(\SU(2))$. Then, the image of the generator $x$ lives on a sphere of the type $S^2_r$ \eqref{2-sphere} or is $\pm\unit$. One easily sees that $\dim \calM =0$ again, but $\dim\zeta = 1$ or $\dim\zeta=3$. This produces a topological part for the divergence degree which is negative. Thus, the highly divergent contribution to the partition function comes from the irreducible representations of the fundamental group.

\section{Conclusion}\label{conclusion}

In this paper, we have sharpened the result of \cite{Bonzom:2010uq} in the context of cellular manifolds $\Delta_M$, by sorting out in the divergence degree  $\Omega$ the topology of $M$ from the cell structure $\Delta_M$. This refined form of $\Omega$ has allowed us to compare it with the other recent results, and to show in which sense our result is more general. In particular, we have emphasized that our geometric approach naturally encompasses the colored model of Gurau \cite{Gurau:2009tw}. We have considered several three-dimensional examples, for which the previous powercounting results are not optimal, to illustrate this point.

We think that an important application of our rewriting of $\Omega$ in terms of topology versus combinatorics is, as we saw in section \ref{jacket-counting}, the insight that it provides into the notion of jacket of colored graphs. The latter indeed allows to probe the fundamental group in a subtle way and leads to a bound on the dimension of the moduli space of flat connections. Hence, it gives an algebraic-topological interpretation of the bounds used by Gurau to perform the $1/N$ expansion, which he obtained in a quite different way. With hindsight, the bound needed for this expansion has two ingredients. The first is the appropriate re-scaling of the coupling constant, which cancels a term proportional to the number of vertices in the graph in the divergence degree. The second involves the genera of jackets to identify the topology of the dominant graphs. Thus, we think our approach could yield an equivalent road to this expansion, clarifying some important aspects.

Our method, in combination with jackets, could also be used to deal with some different models, like the one obtained from the radiative corrections to the Boulatov-Ooguri tensor model in \cite{2ptboulatov} and in which the UV scales are expected to be handled like in the standard model.

\section*{Acknowledgements}
The authors are grateful to Razvan Gurau for so many explanations on his results and also specially for having stressed the importance of the notion of jackets.

\providecommand{\href}[2]{#2}\begingroup\raggedright\endgroup


\pagebreak

\appendix

\section{CW complexes and their (co)homology}

For the reader's convenience, we recall here some basic definitions concerning cell complexes and their (co)homology (a well-written, pedagogical reference is ...). A cell complex, or finite CW complex, is a topological space $X$ presented as the disjoint union of finitely many open cells, such that for each open $p$-cell $\sigma^{p}$ there is a continuous map $f:B^p\rightarrow X$ whose restriction to the interior of $B^p$ is a homeomorphism onto $\sigma^p$, and such that the image of the boundary of $B^p$ is contained in the union of lower dimensional cells. The dimension of $X$, $\dim X$, is the maximal dimension of its cells. If $c_{p}(X)$ is the number of $p$-cells of $X$, the Euler characteristic of $X$ is defined as
\be
\chi(X)=\sum_{p=0}^{\dim X}(-1)^pc_{p}(X).
\ee
It is homotopy invariant.

Cellular homology associates to each cell complex $X$ a sequence of homotopy invariant abelian groups, the homology groups of $X$, as follows. For each dimension $p$, consider the set $C_{p}(X)$ of formal linear combination of $p$-cells with integer coefficients (the `free Abelian group' over the $p$-cells of $X$); its elements are called $p$-chains. Define, for each $p$, the boundary map $\pp_{p}:C_{p}(X)\rightarrow C_{p-1}(X)$ by its action on $p$-cells $\sigma_{\alpha}^p$
\be
\pp_{p}\sigma_{\alpha}^p=\sum_{\beta}[\sigma^{p}_{\alpha},\sigma_{\beta}^{p-1}]\sigma_{\beta}^{p-1}
\ee
and linearity. Here, the sum runs over the $(p-1)$-cells on the boundary of $\sigma^{p}_{\alpha}$, and $[\sigma^{p}_{\alpha},\sigma_{\beta}^{p-1}]$ is the incidence number of $\sigma_{\alpha}^p$ on $\sigma^{p-1}_{\beta}$ -- that is, the number of times $\sigma_{\alpha}^p$ wraps around $\sigma^{p-1}_{\beta}$, with relative orientations taken into account (see ... for a precise definition). Elements of $\ker\pp_{p}$ are called $p$-cycles, and elements of $\textrm{Im}\ \pp_{p+1}$ are called $p$-boundaries. The fundamental property of the boundary maps is that, for each $p$,

\be
\pp_{p}\pp_{p+1}=0.
\ee

This implies that $\textrm{Im}\ \pp_{p+1}\subset\ker\pp_{p}$, and allows to consider the quotients $H_{p}(X)=\ker\pp_{p}/\textrm{Im}\ \pp_{p+1}$ of $p$-cycles modulo $p$-boundaries. The sequence of Abelian groups $C_{p}(X)$, together with the boundaries maps $\pp_{p}$
\be
0\longrightarrow C_{\dim X}(X)\longrightarrow\dots\longrightarrow C_{p+1}(X)\longrightarrow C_{p}(X)\longrightarrow \dots\longrightarrow C_{0}(X)\longrightarrow 0
\ee
forms the cellular chain complex of $X$, and the $H_{p}(X)$'s are the homology groups of $X$. They are homotopy invariant, and in particular so are their ranks $b_{p}(X)$, the Betti numbers of $X$. Intuitively, $b_{p}(X)$ is the number of `independent $p$-holes' of $X$.
The Euler-Poincar\'e theorem states that
\be
\chi(X)=\sum_{p=0}^{\dim X}(-1)^pb_{p}(X).
\ee

Dualization of this construction leads to cellular cohomology. Explicitely, for each p, the cochain group $C^p(X)$ of $X$ is defined as the set of linear maps from $C_{p}(X)$ to $\mathbb{Z}$, and the coboundary operator $\delta^{p}$ as the transpose of $\pp^{p+1}$. One checks that
\be
\delta^{p+1}\delta^{p}=0
\ee
and the resulting complex
\be
0\longrightarrow C_{0}(X)\longrightarrow\dots\longrightarrow C_{p}(X)\longrightarrow C_{p+1}(X)\longrightarrow \dots\longrightarrow C_{\dim X}(X)\longrightarrow 0
\ee
is called the cellular cochain complex of $X$, and the quotients $H^p(X)=\ker\delta_{p}/\textrm{Im}\ \delta_{p-1}$ its cohomology groups. They are homotopy invariant as well.

This construction can be generalized by replacing the integer coefficients in the (co)chains by elements of an arbitrary Abelian group $A$. The corresponding homology and cohomology groups are then denoted $H_{p}(X,A)$ and $H^p(X,A)$ respectively. One shows in particular that, whenever $A$ is actually a vector space, $H_{p}(X,A)$ and $H^p(X,A)$ are dual to each other. Also, in this case (and more generally if $A$ is torsion-free), it holds that $H_{p}(X,A)=H_{p}(X)\otimes A$, and thus $b_{p}(X,A)=b_{p}(X)\dim(A)$.

Eventually, let us mention Poincar\'e duality: when $X$ is the cell decomposition of an oriented, closed $d$-manifold, we have

\be
H^p(X)\simeq H_{d-p}(X).
\ee

\section{Relationship between colored and cellular homology}\label{colored}

Due to the several simplifications it provides, the recent literature on group field theory has focused on Gurau's colored model \cite{Gurau:2009tw}. To ease the translation between this framework and ours, based on finite CW complexes, we recall here the basics of cristallization and colored graph theory, and explicit the relationship between colored and cellular homology. Although it seems that Gurau did not know about this work, colored homology was actually introduced in \cite{Cavicchioli199599}.

An $(d + 1)$-colored graph is a pair $\mathcal{G}=(G, c)$, where $G = (V(G), E(G))$ is a connected multigraph (without tadpoles) regular of degree $d + 1$, and $c : E(G)\rightarrow\Delta_d = \{0, 1, . . . , d\}$ is an edge-coloring such that $c(e)\neq c(f )$ for any pair $e$ and $f$ of adjacent edges of $G$. For each proper subset $B$ of $\Delta_n$, let $\mathcal{G}_B$ denote the subgraph of $G$ defined by $(V (G), c^{-1}(B))$; each connected component of $\mathcal{G}_B$ is called a $B$-residue, or $B$-bubble. If $B$ has $\vert B\vert=k$ elements, it is also called a $k$-bubble. The $0$-bubbles of $\mathcal{G}$ are its vertices.

The \emph{colored}, or \emph{bubble}, homology of such a colored graph $\mathcal{G}$ is defined as follows. Let $C_k(\mathcal{G})$ be the free Abelian group generated by the $k$-bubbles of $\mathcal{G}$, and $d_k:C_k(\mathcal{G})\rightarrow C_{k-1}(\mathcal{G})$ the linear map defined by
\be\label{coloredboundary}
d_k(b_k):=\sum_{q\in B}(-1)^{\vert B^<(q)\vert}\sum_{c_{k-1}\in\mathcal{G}_{B\setminus\{q\}}}c_{k-1}
\ee
where $b_k$ is a $B$-bubble $k$ colors, and
\be
B^<(q):=\left\{p\in B\ ,\ p<q\right\}.
\ee
It is not difficult to check that $d_{k}d_{k+1}=0$, and thus that
\be
C_k(\mathcal{G})\xrightarrow{\ d_d\ }
C_{d-1}(\mathcal{G})\xrightarrow{\ d_{d-1}\ }  \dots C_1(\mathcal{G})\xrightarrow{\ d_1\ }C_0(\mathcal{G})\xrightarrow{\ d_0\ }0,
\ee
forms a chain complex. Its homology groups $H_k(\mathcal{G})=\ker d_k/\im d_{k+1}$ are the \emph{colored homology groups} of $\mathcal{G}$. We denote the corresponding Betti numbers $b_k(\mathcal{G})$.

\medskip

Every $(d + 1)$-colored graph $(G, c)$ determines a $d$-dimensional CW complex $K_{\mathcal{G}}$ as follows. For each vertex $v$ of $G$, consider an $d$-simplex $\sigma_d(v)$ and label its vertices by $\Delta^d$. If $v$ and $w$ are joined in $G$ by an $i$-colored edge, $i\in\Delta_d$, then identify the $(d - 1)$-faces of $\sigma_d(v)$ and $\sigma_d(w)$ opposite to the vertex labelled by $i$, so that equally labelled vertices coincide. The quotient space is $K_{\mathcal{G}}$. It is obviously connected.

The CW complex thus defined is special in that every cell $e^k_\alpha$ arises as the projection of a $k$-simplex $\sigma^k_\alpha$ with vertices labelled in a subset $B_\alpha$ of $\Delta^d$. Moreover, the cells on the boundary of $e^k_\alpha$ are the projections of the faces of $\sigma^k_\alpha$. Such a CW complexes is called a $\Delta$-complex by Hatcher \cite{hatcher}, and a pseudo-complex by other authors \cite{Cavicchioli199599}. Thanks to its simplicial character, its boundary operator is readily computed.  For $q\in B_\alpha$, denote $e^{k-1}_{\alpha(q)}$ the $(k-1)$-cell on the boundary of $e^k_\alpha$ arising as the projection of the face of $\sigma^k_\alpha$ opposite to the vertex labelled by $q$. Then $[e^k_\alpha,e^{k-1}_{\alpha(q)}]=(-1)^{\vert B_\alpha^<(q)\vert}$, and thus
\be
\pp_k(e^k_\alpha)=\sum_{q\in B_\alpha}(-1)^{\vert B_\alpha^<(q)\vert}e^{k-1}_{\alpha(q)}.
\ee

Moreover, like a simplicial complex, $K_{\mathcal{G}}$ possesses a dual CW complex $\Delta_{\mathcal{G}}$, whose $k$-cells $f^k_\alpha$ are in one-to-one correspondence with the $(d-k)$-cells $e^{d-k}_\alpha$ of $K_{\mathcal{G}}$ (see \cite{hatcher} for details). Let
\be
E^k_{\alpha,q}=\left\{\beta\ ,\ e^{k}_{\beta(q)}=e^k_\alpha\right\}
\ee
index the set of cofaces of $e^k_\alpha$. Then $\Delta_{\mathcal{G}}$ is such that
\be\label{cellularboundary}
\pp_k(f^k_\alpha)=\sum_{q\in B_\alpha}(-1)^{\vert B_\alpha^<(q)\vert}\sum_{\beta\in E_{\alpha,q}^k} f^{k-1}_\beta,
\ee
i.e. the boundary operator $\pp_k$ of $\Delta_{\mathcal{G}}$ is the coboundary operator $\delta^{d-k}$ of $K_{\mathcal{G}}$.  Hence, \be H_k(\Delta_{\mathcal{G}})=H^{d-k}(K_{\mathcal{G}}),\ee as in Poincar\'e duality.

To relate the colored homology of the colored graph $\mathcal{G}$ to the cellular homology of the corresponding complex, it suffices to note that the $B$-bubbles of $\mathcal{G}$ are in one-to-one correspondence with the cells of $K_{\mathcal{G}}$ arising from simplices with vertices labelled in $\Delta^d\setminus B$. Indeed, let $v$ be a vertex of a $B$-bubble $b$ with $\vert B\vert=k$, and $\sigma_d(v)$ the corresponding simplex, with vertices labelled by $\Delta_d$. Let $f_B(\sigma_d(v))$ be the $(d-k)$-subsimplex of $\sigma_d(v)$ defined by those of its vertices which are not in $B$, and $e^{d-k}_b(v)$ the corresponding $(d-k)$-cell in $K_{\mathcal{G}}$. By definition of the quotient space $K_{\mathcal{G}}$, the cell $e^{d-k}_b(v)$ actually does not depend on $v$. Thus, for each $k$-bubble $b$ of $\mathcal{G}$ there is a $(d-k)$-cell $e^{d-k}_b$ of $K_{\mathcal{G}}$, and therefore a $k$-cell $f^k_b$ of $\Delta_{\mathcal{G}}$. It is not hard to check that this correspondence is actually bijective. Moreover, inspection of \eqref{coloredboundary} and \eqref{cellularboundary} shows that
\be
F_k(d_kb)=\delta^{d-k}F_k(b),
\ee
where $F_k$ the mapping $b\mapsto f^k_b$. In other words, we have an isomorphism of chain complexes $F=(F_k)_{k=0}^d:C_*(\mathcal{G})\rightarrow C^{*}(\Delta_{\mathcal{G}})$, and thus
\be
H_k(\mathcal{G})\simeq H_k(\Delta_{\mathcal{G}})=H^{d-k}(K_{\mathcal{G}}).
\ee
In particular, the colored Betti numbers of $\mathcal{G}$ coincide with the cellular Betti numbers of $\Delta_{\mathcal{G}}$. Note moreover that, since the $0$-th Betti number of a topological space counts the number of its connected components, we have $b_d(\mathcal{G})=b_0(K_{\mathcal{G}})=1$ and $b_0(\mathcal{G})=b_0(\Delta_{\mathcal{G}})=1$.

\end{document}